\documentstyle[amsfonts,aps,prl,twocolumn,epsf]{revtex}

\newcommand{\beq}{\begin{equation}}
\newcommand{\eeq}{\end{equation}}
\newcommand{\beqa}{\begin{eqnarray}}
\newcommand{\eeqa}{\end{eqnarray}}

\begin{document}

\title{Irrelevance of memory in the minority game}

\author{
Andrea Cavagna\thanks{E-mail: a.cavagna1@physics.ox.ac.uk}}

\address{
Theoretical Physics, University of Oxford,
1 Keble Road, Oxford, OX1 3NP, UK
\vskip 0.5 truecm }

\date{\today}
\maketitle

\begin{abstract}
{{\bf Abstract:} 
\sl By means of extensive numerical simulations we show that all
the distinctive features of the minority game introduced by
Challet and Zhang (1997), are completely independent from the 
memory of the agents. 
The only crucial requirement is that all the individuals must 
posses the same information, irrespective of the fact that this 
information is true or false.} 
\end{abstract}

\vskip 0.5 truecm

Originally inspired by the {\it El Farol} problem stated by 
Arthur in \cite{arthur}, it has been introduced in \cite{challet1} 
a model system for the adaptive evolution of a population 
of interacting agents, the so called minority game. 
This is a toy model where inductive, rather than deductive, 
thinking, in a population of bounded rationality, gives rise to 
cooperative phenomena.

The setup of the minority game is the following: 
$N$ agents have to choose 
at each time step whether to go in room $0$ or $1$. 
Those agents who have chosen the less crowded room 
(minority room) win, the other loose, so that the system 
is intrinsically frustrated.

A crucial feature of the model is the way by which agents 
choose. In order to decide in what room to go, agents use 
strategies. 
A strategy is a choosing device, that is an object that 
processes the outcomes of the winning room in the last $m$
time steps (each outcome being $0$ or $1$) and accordingly 
to this information prescribes in what room to go the 
next step. 
The so-called memory  $m$ defines $2^m$ potential past 
histories (for instance, with $m=2$ there are four possible 
pasts, $11$, $10$, $01$ and $00$). A strategy is thus formally a 
vector $R_{\mu}$, with $\mu=1,\dots,2^m$, whose elements 
can be $0$ or $1$. The space $\Gamma$ of the strategies is 
an hypercube of dimension $D=2^m$ and the total number of 
strategies is $2^D$.

At the beginning of the game each agent draws randomly a 
number $s$ of strategies from the space $\Gamma$ and 
keeps them forever, as a genetic heritage.
The problem is now to fix which one, among these $s$
strategies, the agent is going to use
\footnote{We will consider only the non-trivial case $s>1$.}.
The rule is the following.
During the game the agent gives point to {\it all} his/her \
strategies according to their potential success: at each 
time step a strategy gets a point only if it has forecast the 
correct winning room, regardless of having been actually 
used or not. At a given time the agent chooses among 
his/her $s$ strategies the most successful one up to that 
moment (i.e. the one with the highest number of points) and 
uses it in order to choose the room. The adaptive nature
of the game relies in the time evolution of the best strategy
of each single agent.

In this way the game has a well defined deterministic time 
evolution, which only depends on the initial distribution
of strategies and on the random initial string of $m$ bits 
necessary to start the game.

Among all the possible observables, a special role is played by
the variance $\sigma$ of the attendance $A$ in a given room 
\cite{challet1}. We can consider, for instance, room 
$0$ and define $A(t)$ as the number of agents in this room 
at time $t$. We have,
\beq
	\sigma^2=\lim_{t\to\infty}\frac{1}{t} 
	\int_{t_0}^t  dt' \; \left( A(t')-\frac{N}{2} \right)^2  \ ,
\eeq
where $N/2$ is the average attendance in the room and $t_0$ 
is a transient time after which the process is
stationary \cite{challet1,savit1}.
In all the simulations presented in this Letter it has been  
taken $t=t_0=10,000$ for a maximum value of $N=101$ and it 
has been verified that the averages were saturated over 
these times.

The importance of $\sigma$ (called {\it volatility} in financial 
context) is simple to understand: the larger is $\sigma$, the 
larger is the global waste of resources by the community of agents.
Indeed, only with an attendance $A$ as near as possible to its
average value there is the maximum distribution of points 
to the whole population.
Moreover, from a financial point of view, it is clear that a low
volatility $\sigma$ is of great importance in order to minimize 
the risk.  

If all the agents were choosing randomly, the va\-rian\-ce would
simply be $\sigma_r^2=N/4$.
An important issue is therefore: in what conditions is the 
variance $\sigma$ {\it smaller} than $\sigma_r$ ?
In other words, is it possible for a population of selfish 
individuals to collectively behave in a better-than-random way ?
What has been found first in \cite{savit1} is that the volatility
$\sigma$ as a function of $m$ has a remarkable 
behaviour, since actually {\it there is} a regime 
where $\sigma$ is smaller than the random value $\sigma_r$. 
In this phase the collective behaviour is such that
less resources are globally wasted by the population of agents.
A deep understanding of this feature is therefore important.

From the very definition of the model and from the behaviour of
$\sigma(m)$ described above, it seems clear that the memory 
$m$ is a crucial quantity for the two following reasons.

First, from a geometrical point of view, $m$ defines the dimension 
of the space of strategies $\Gamma$ and therefore it is related to 
the probability that strategies drawn randomly by different agents
could give similar predictions: the larger is $m$, the bigger 
is $\Gamma$ and the lower is the pro\-ba\-bi\-li\-ty that different 
players have some strategies in common. 
Since the non-random nature of the game relies in the presence 
of correlated choices, that is, exactly in the possibility that 
different agents use the same strategies, it follows that for 
very large $m$ the game proceeds in a random way
\cite{savit1,challet2,savit2,johnson3}
\footnote{This argument works at fixed number of agents $N$. 
O\-ther\-wise the relevant variable will be $2^m/N$. We discuss
this point later.}. 

Secondly, $m$ is supposed to be a real memory. A\-ctual\-ly, 
the whole game is constructed around the role of $m$ as
a memory: at time $t$ agents use strategies which process 
the last $m$ events in the past. As a consequence of this, 
a new minority room will come out and at time $t+1$ there 
will be a new $m$-bits past which will differ from the old 
one for the outcome at time $t$. 
Thus, agents, or better, strategies, choose by remembering 
the last $m$ steps of time history, so that $m$ is a natural
time scale of the system. 
Due to this, an explanation of the behaviour of $\sigma(m)$
has been proposed in \cite{savit1}, where the decay 
rate of the time correlations in the system is compared and 
related to $m$, thus supporting the key interpretation 
of $m$ as a real memory.
This memory role of $m$ complicates greatly the nature 
of the problem, since it induces an explicit dynamical 
feedback in the evolution of the system, such that the 
process is not local in time.


The purpose of this Letter is to show that the memory 
of the agents is irrelevant. We shall prove that there 
is no need of an explicit time feedback, to obtain all 
the distinctive features of the model. 


In order to prove this statement we consider the same model 
introduced in \cite{challet1} and described above, 
but with the following important difference: 
at each time step, the past history is just {\it invented}, 
that is, a random sequence of $m$ bits is drawn, to play 
the role of a fake time history. 
This is the information that all the agents process with 
their best strategies to choose the room.
As we are going to show, this oblivious version 
of the model gives exactly the same results as the 
original one, thus proving that the role of $m$ is 
purely geometrical. 

In Fig.1, the variance $\sigma$ as a function of $m$ is plotted both
for the case with and without memory. The two models give the same 
results, not only qualitatively, but also quantitatively (see also
the data of \cite{savit1,challet2,savit2,johnson3}). 
In par\-ti\-cu\-lar, the minimum of $\sigma$ as a function of $m$ is 
found even without memory and cannot therefore be related to it.
\begin{figure}
\begin{center}
\leavevmode
\epsfxsize=3in
\epsffile{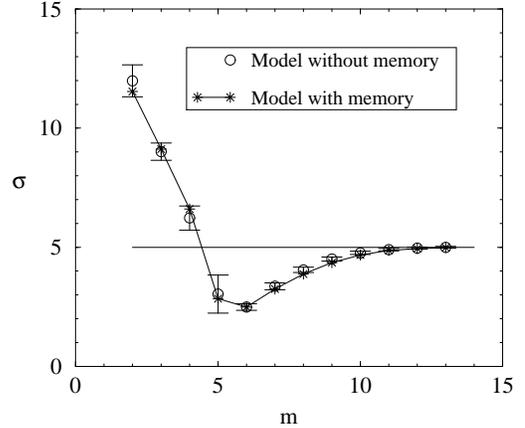}
\caption{Model without memory vs. model with memory. 
The variance $\sigma$ as a function of $m$ for $s=2$. 
The horizontal line is the variance $\sigma_r$ of the random case.
The number of agents is $N=101$. Average over $100$ samples.
Errors bars are shown only for the model without memory, while 
the line just connects the points of the memory model.}
\label{fig1}
\end{center}
\end{figure}

The dependence of the whole function $\sigma(m)$ on the 
individual number of strategies $s$ is another important point. 
It has been shown for the first time in \cite{challet2} that the 
minimum of this curve is shallower the larger is the value of $s$. 
In Fig.2 we show that this same phenomenon occurs for the model 
without memory. 
\begin{figure}
\begin{center}
\leavevmode
\epsfxsize=3in
\epsffile{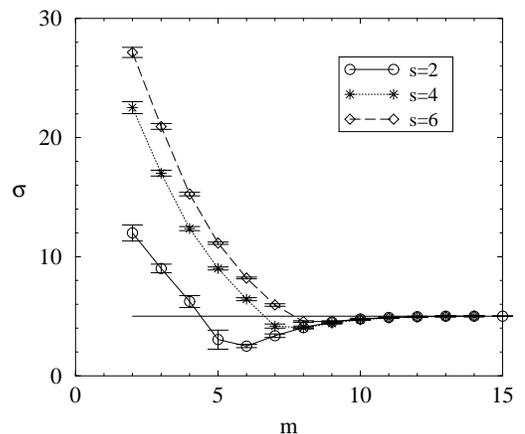}
\caption{Model without memory. Variance $\sigma$ as a
function of $m$, at different values of $s$, $N=101$. Average over
$100$ samples. Lines are just a guide for the eye.}
\label{fig2}
\end{center}
\end{figure}

From a technical point of view, note that, once eliminated 
the role of $m$ as a memory, the only quantity involved in the 
actual implementation of the model is $D$, the dimension of the 
space of strategies $\Gamma$. Therefore, instead of drawing a random
sequence of $m$ bits, it is much easier to draw a random 
component $\mu\in[1,D]$ to mimic the past history:
each agent uses component $\mu$ of his/her\ best strategy 
to choose the room. The main consequence of this is that
there is no need for being $D=2^m$, since we can choose any integer 
value of $D$. In \cite{savit2} it has been introduced a method
by which it is possible to consider non-integer values of $m$ in
the model with memory. This is useful, since it permits to study the
shape of $\sigma(m)$ around its minimum, with a better resolution 
in $m$. In the present context, it is trivial to consider non-integer 
values of $m$, since we simply have $m=\log_2 D$. In this way 
results identical to \cite{savit2} are obtained.

Once fixed $s$, let $m_c$ be the value of $m$ where the 
minimum of $\sigma(m)$ occurs. 
In \cite{savit1} it has been pointed out that for $m<m_c$ the 
variance $\sigma$ grows as $N$, where $N$ is the number of agents,
while for $m>m_c$ it grows as $N^{1/2}$. In Fig.3, $\sigma$ as
a function of $N$ is plotted for the model without memory. 
The same behaviour as in the model with memory is found.
\begin{figure}
\begin{center}
\leavevmode
\epsfxsize=3in
\epsffile{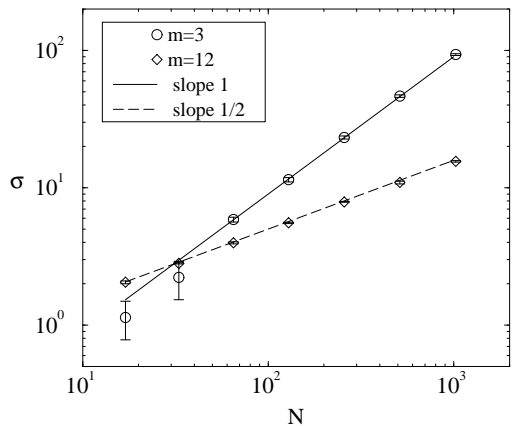}
\caption{Model without memory. Variance $\sigma$ as a function of
the number of agents $N$, for different values of $m$, at fixed $s=2$.
Average over $10$ samples. The full line is $\sigma\sim N$, while
the dashed line is $\sigma\sim N^{1/2}$.} 
\label{fig3}
\end{center}
\end{figure}

An interesting question is whether $\sigma$ is a function of 
a single scaling variable $z$ constructed with $m$, $N$ and $s$.
It has been shown in \cite{savit1} that by considering as a scaling 
variable $z=2^m/N=D/N$
all the data for $\sigma$ at various $m$ and $N$ collapse on the 
same curve. In this case the relevant parameter is thus 
the dimension $D$ of $\Gamma$, over the number $N$ of 
playing strategies.
On the other hand, it has been proposed in \cite{challet2} a 
different scaling variable, that is $z'=2\cdot 2^m/sN=2D/sN$. 
In this way, the relevant parameter would be the density on 
$\Gamma$ of the {\it total} number of strategies 
$sN$. In Fig.4 we plot $\sigma^2/N$ as a function of $z'$, at  
different values of $D$, $N$ and $s$, for the model without 
memory. We see that the correct scaling parameter is $z$ and not 
$z'$, since the data with different values of $s$ collapse on 
different curves. The same result is obtained if we perform the 
simulation with the memory (see \cite{savit2}). 
The two models give once again the same results.
Note from Fig.4 that the scaling is not perfect at very low
values of $z'$, that is for very small $D$. This is just a
trace of the integer nature of the model.

\begin{figure}
\begin{center}
\leavevmode
\epsfxsize=3in
\epsffile{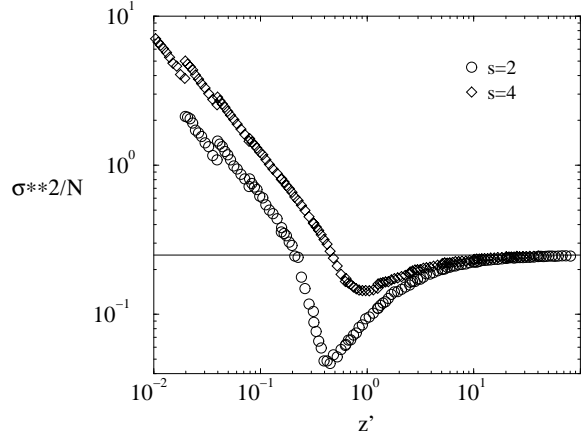}
\caption{Model without memory. Plot of $\sigma^2/N$ as a
function of the scaling parameter $z'=2D/sN$. 
The number of agents $N$ varies from $N=51$ to
$N=101$, while $D$ varies from $D=2$ to $D=4096$.
The individual number of strategies $s$ ranges over two 
values $s=2$ and $s=4$. Average over $50$ samples.}
\label{fig4}
\end{center}
\end{figure}

From what shown above it is reasonable to conclude 
that, in order to obtain all the crucial features of 
the minority game, the presence of an individual 
memory of the agents is irrelevant. 
The parameter $m$ still plays a major role, but only 
for being related to the dimension $D=2^m$ of the strategies 
space $\Gamma$. 
A consequence of this fact is that any attempt to 
explain the properties of this model, relying on the role 
of $m$ as a memory, can hardly be correct.
On the other hand, as already said, the geometrical role
of $m$ remains. Indeed, some recent attempts to give
an analytic description of the model (see 
\cite{challet2,johnson3}) are only grounded on 
geometrical considerations about the distribution of 
strategies in the space $\Gamma$ and go therefore, in our 
opinion, in the correct direction.

The most important result of the present Letter is the existence
of a regime where the whole population of agents still behaves 
in a better-than-random way, even if the information 
they process is completely random, that is wrong, if compared 
to the real time history.
{\it The crucial thing is that everyone must possess the same
information}. Indeed, if we invent a different past 
history for each different agent, no coordination emerges at all
and the results are the same as if the agents were behaving 
randomly (this can be easily verified numerically). 
In other words, if each individual is processing a different 
information, the features of the system are completely identical 
to the random case, irrespective of the values of $m$ and $s$. 

The conclusion is the following: the crucial property is not 
at all the agents' memory of the real time history, but rather the 
fact that they all share the same information, whatever false 
or true this is. As a consequence, there is no room in this 
model for any kind of forecasting of the future based on the 
``understanding'' of the past.

We hope this result to be useful for a future deeper 
understanding of this kind of adaptive systems.
Indeed, before trying to explain the rich structure of a 
quite complicated model, it is important in our opinion 
to clear up what are the truly necessary ingredients of 
such a model and what, on the contrary, is just an irrelevant 
complication, which can be dropped.
In the case of the so-called memory (or brain size, or 
intelligence), $m$, there also has been a problem of 
terminology: given the original formulation of the model, 
it seemed that the very nature of a variable encoding 
the {\it memory} or the {\it intelligence} of the agents, 
could warrant by itself a relevance to it 
\cite{challet1,savit1,challet2,savit2,johnson3,johnson1,johnson2}, 
relevance which, 
as we have seen, was not deserved. 
Notwithstanding this, we consider the present model still 
to be very interesting and far from being trivial.

Finally, let us note that the passage from a model with memory 
to a model without memory, is equivalent to substitute a 
deterministic, but very complicated system, with a stochastic, 
but much simpler one, which nevertheless gives the same results 
as the original case and which is therefore indistinguishable 
from it for all the practical purposes.
The use of a stochastic/disordered model to mimic a 
deterministic/ordered one, is similar in the spirit to 
what happens in the context of glassy systems, where some 
disordered models of spin glasses are often used in order 
to have a better understanding of structural glasses, 
which contain in principle no quenched disorder 
\cite{mezard-parisi-glass}.

\acknowledgements
I wish to thank Erik Aurell, Francesco Bucci, 
Juan P. Garrahan, John Hertz and David Sherrington 
for useful discussions and in particular Irene Giardina 
for many suggestions and for reading the manuscript. 
I also thank for the kind hospitality NORDITA (Copenhagen), 
where part of this work has been done. 
This work is supported by EPSRC Grant GR/K97783.  

\end{document}